\documentclass[twocolumn,pra,showpacs,preprintnumbers,amsfonts,amsmath,amssymb,superscriptaddress]{revtex4}

\usepackage{graphicx}
\usepackage{dcolumn}
\usepackage{bm}
\usepackage[bookmarks=false,pdfstartview={FitH}]{hyperref}

\newcommand{\ba}{\begin{eqnarray}}
\newcommand{\ea}{\end{eqnarray}}
\newcommand{\be}{\begin{equation}}
\newcommand{\ee}{\end{equation}}
\newcommand{\bea}{\begin{eqnarray}}
\newcommand{\eea}{\end{eqnarray}}

\usepackage{theorem}

\theorembodyfont{\rmfamily}

\theoremstyle{break}
\def\QED{~\rule[-1pt]{5pt}{5pt}\par\medskip}

\begin{document}


\title{Elliptic functions and efficient control of Ising spin chains with unequal couplings}

\author{Haidong Yuan}%
\email{haidong@mit.edu}
\affiliation{%
Department of Mechanical Engineering,
Massachusetts Institute of Technology,
77 Massachusetts Avenue,
Cambridge, Massachusetts 02139, USA
}%
\author{Robert Zeier}%
\email{zeier@eecs.harvard.edu}
\affiliation{%
Harvard School of Engineering and Applied Sciences,
33 Oxford Street,
Cambridge, Massachusetts 02138, USA
}%
\author{Navin Khaneja}%
\email{navin@hrl.harvard.edu}
\affiliation{%
Harvard School of Engineering and Applied Sciences,
33 Oxford Street,
Cambridge, Massachusetts 02138, USA
}%

\date{\today}

\begin{abstract}
In this article, we study optimal control of dynamics in
a linear chain of three spin $1/2$, weakly coupled with 
unequal Ising couplings. We address the problem of 
time-optimal synthesis of 
multiple spin quantum coherences. 
We derive time-optimal pulse sequence for creating a desired 
spin order by computing geodesics on a sphere under a special metric. 
The solution to the geodesic equation is 
related to the nonlinear oscillator equation and the 
minimum time to create multiple spin
order can be expressed in terms of an elliptic 
integral. These techniques are used for efficient 
creation of multiple spin coherences in Ising spin-chains 
with unequal couplings.

\end{abstract}


\maketitle

\section{\label{sec:introduction}Introduction}

In the absence of relaxation, experiments in coherent spectroscopy
and quantum information processing consist of a 
sequence of unitary transformations on the quantum system of interest. 
Pulse sequences that create a desired unitary transformation in the minimum 
possible time are of particular interest in experimental realizations as 
they can reduce losses due to relaxation. This poses the problem of 
time optimal control of quantum systems, which is of both theoretical 
and practical interest in the broad area of coherent control of quantum systems.

Besides application in spectroscopy, synthesizing unitary transformations 
in a time-optimal way using available physical resources is also a problem in quantum 
information processing. It has received
significant attention, and time-optimal control of two coupled qubits
\cite{KBG:2001, BCL:2002, VHC:2002, HVC:2002, RKG:2002, ZGB:2004, YK:2005, Yua:2006, Zei:2006} 
is now well understood. Recently, this problem has also been studied in the
context of linearly coupled three-qubit topologies
\cite{KGB:2002, RKG:2003, KHSY:2007}, where significant savings in the implementation
time of trilinear Hamiltonians and logic gates between indirectly coupled qubits 
were demonstrated over conventional methods. Many of these ideas have found 
applications in efficient ways to propagate coherences along Ising spin chains 
\cite{KG:2002, YGK:2007}.
However, the complexity of the general problem of time-optimal control of multiple 
qubit topologies is only beginning to be appreciated.

In this article, we study the problem of finding the shortest pulse sequences
for creating desired coherences 
in a linear chain  of spins (of spin $1/2$) weakly coupled with unequal Ising couplings.
In particular, we start by considering the case of three linearly coupled 
spins and generalize
our methods to linear spin chains. This study has immediate applications to 
multi-dimensional nuclear magnetic resonance (NMR) 
spectroscopy \cite{EBW:1997}. In multi-dimensional NMR experiments \cite{EBW:1997}, 
starting from the thermal state of the spins,
a multiple quantum coherence between spins is synthesized, which helps to 
correlate the frequencies of various spins. Efficient pulse sequences for creating 
such coherences help to improve the sensitivity of the experiments.

One approach for solving these problems  reduces the efficient 
synthesis of multiple spin order to the
geometrical question of finding shortest paths on the sphere under 
a special metric \cite{KBG:2001,Yua:2006,YK:2006,KHSY:2007}.
The metric enforces the constraints on the quantum
dynamics that arise because only limited Hamiltonians can be realized 
~\cite{NDGD:2006,Nie:2006,NDGD:2006b}.
 Such analogies between optimization problems related to steering dynamical 
systems with constraints and geometry have been well explored in areas of control
theory \cite{Bro:1982, Bal:1975} and sub-Riemannian geometry
\cite{Mon:2002}.  In this paper, we study in detail the metric and
the geodesics that arise from the problem of efficient
synthesis of multiple spin coherences between qubits (spin $1/2$) 
that are indirectly coupled via unequal couplings to a third qubit.    

\section{Optimal Control of a Linear Three-Spin System with Unequal Couplings}
We consider a linear chain of three spins placed in a static
external magnetic field in the $z$ direction with Ising type
couplings between next neighbors \cite{Isi:1925, Cas:1989}. In a
suitably chosen (multiple) rotating frame which rotates with each
spin at its resonant frequency, the Hamiltonian that governs the
free evolution of the spin system is given by the 
coupling Hamiltonian
$$ H_c=2 J_{12}I_{1z}I_{2z}+ 2  J_{23}I_{2z}I_{3z}. $$
We use the notation 
$I_{\ell \nu}= \bigotimes_{j} I_{a_{j}}$, where $a_{j}=\nu$ for $j=\ell$ and $a_{j}=0$ otherwise  (see \cite{EBW:1997}).
The matrices 
$I_{x}:=
\left(\begin{smallmatrix}
0 & 1 \\
1 & 0
\end{smallmatrix}
\right)/2$,
$
I_{y}:=
\left(
\begin{smallmatrix}
0 & -i \\
i & 0
\end{smallmatrix}
\right)/2
$, and
$
I_{z}:=
\left(
\begin{smallmatrix}
1 & 0 \\
0 & -1
\end{smallmatrix}
\right)/2
$, are the Pauli spin matrices
and
$
I_{0}:=
\left(
\begin{smallmatrix}
1 & 0 \\
0 & 1
\end{smallmatrix}
\right)
$
is the $2\times 2$-dimensional identity matrix.

If the Larmor frequencies of the spins are well separated, each
spin can be selectively excited by an appropriate choice of the
amplitude and phase of the radio-frequency (RF) field at its resonance frequency. The
goal of the pulse designer is to derive 
explicit controls for the variables comprising of the frequency, amplitude and phase
of the external RF field to effect a net unitary evolution $U(t)$
most efficiently.

We begin with the problem of finding the shortest pulse sequence
that transform the initial polarization $I_{1z}$ on the first spin
to a multiple quantum coherence, i.e.,
$$ I_{1x} \rightarrow 4I_{1z}I_{2z}I_{3z}. $$

{\it Example 1.}
The conventional strategy for achieving this transfer is given by 
the following stages
\begin{eqnarray*}
I_{1x} \stackrel{(H_c)_{\tau_1}}\rightarrow 2I_{1y}I_{2z} \stackrel{I_{2y}}\rightarrow 2I_{1y}I_{2x} \stackrel{(H_c)_{\tau_2}} \rightarrow 4I_{1y}I_{2y}I_{3z}
\end{eqnarray*}
In the first stage of the transfer, operator $I_{1x}$ evolves to
$2I_{1y}I_{2z}$ under the natural coupling Hamiltonian 
$H_c$ in $\tau_1 = \pi/(2J_{12})$ units of time. This operator is then
rotated to $2I_{1y}I_{2x}$ by applying a hard $(\pi/2)_y$ pulse 
on the second spin, which evolves to $4I_{1y}I_{2y}I_{3z}$ under the 
natural coupling, in $\tau_2 = \pi/(2J_{13})$ units of time. 
Finally, hard  $(\pi/2)_x$ pulses on first and second spin prepare 
the desired final state. The total evolution time is then
simply $\tau_1+\tau_2 = \pi/(2J_{12}) + \pi/(2J_{13})$.

We now study time-optimal designs for achieving this (and more general) transfers.
To simplify notation, we introduce the following symbols $\langle O\rangle:=\mathrm{Tr}(O \rho)$, where $\mathrm{Tr}$ denotes the trace, for the
expectation values of operators $O$. Let 
$x_1=\langle I_{1x} \rangle$, 
$x_2=\langle 2I_{1y}I_{2z} \rangle$,
$x_3=\langle 2I_{1y}I_{2x} \rangle$, and
$x_4=\langle 4I_{1y}I_{2y}I_{3z} \rangle$ and $X = (x_1, x_2, x_3, x_4)^T$.
By expressing the time $t$, in units of $1/J_{12}$, the evolution of the vector $X$ is
given by
\begin{equation}
\label{eq:transfer} \frac{dX}{dt}= 
    \begin{pmatrix}
      0 & -1 & 0 & 0 \\
      1 & 0 & -u & 0 \\
      0 & u & 0 & -k \\
      0 & 0 & k & 0 
      \end{pmatrix} X    =  B(u, k) X,
\end{equation}
where 
$k = J_{23}/J_{12}$
and $u=u(t)$ is the control parameter
representing the amplitude of the 
$y$-pulse on the second spin. Now, the problem of optimal transfer
is to find the optimal
$u(t)$ for steering the system from $(1,0,0,0)^{T}$ to 
$(0,0,0,1)^{T}$ in  minimal
time. We consider also the more general case of transfering
the system from
$(\cos\alpha,\sin\alpha,0, 0)^{T}$ to $(0, 0, \cos\beta,\sin\beta )^{T}$, where
we consider different $\alpha,\beta\in [0,\pi/2]$.

{\it Example 1\, (continued).}
In this picture, the conventional method to transfer
$(1,0,0,0)^{T}$ to $(0,0,0,1)^{T}$
is described by setting 
$u(t)=0$ for $\tau_1$ units of time, 
transfering $(1,0,0,0)^{T}$ to $(0,1,0,0)^{T}$. Using the control
$u$, we rotate $(0,1,0,0)^{T}$ to $(0,0,1,0)^{T}$ in arbitrary small time, as the control can 
be performed much faster as compared to the evolution of couplings.
Then, we set $u(t)=0$ and evolve $\tau_2$ units of time under the coupling 
Hamiltonian, transfering $(0,0,1,0)^{T}$ to $(0,0,0,1)^{T}$. The total 
time for the transfer is $\tau_1 +\tau_2$.

We now show how significantly shorter transfer times are achievable, if we 
relax the constraint that the selective rotations on second spin are only 
hard pulses. We show that if we let the selective operations be carried out by 
soft shaped pulses, along with evolution of the coupling Hamiltonian, then we can 
achieve shorter transfer times. The pulse shapes can be numerically computed by 
formulating the problem as a derivation of geodesics on a sphere 
under a special metric as detailed below.

\begin{figure}
\includegraphics{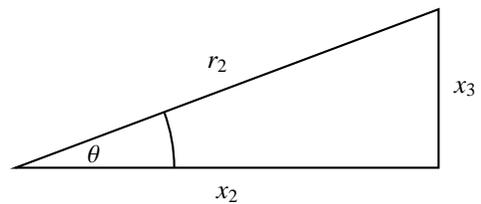}
\caption{Auxiliary variables $r_2$ and $\theta$.\label{fig:1}}
\end{figure}

We first make a change of variables (see Fig.~\ref{fig:1}). Let
$$r_1=x_1, r_2=\sqrt{x_2^2+x_3^2}, r_3=x_4,\, \text{and\,} \tan \theta=\frac{x_3}{x_2}.$$
Using $u(t)$, we can control the angle $\theta$, so we can think of
$\theta$ as a control variable. Expressing the time in units
of $1/J_{12}$, the evolution of the system w.r.t.\ the coordinates
$r_i$ is given by
\begin{equation}\label{eq:main0}
\frac{d}{dt}\begin{pmatrix}
           r_1\\
           r_2\\
           r_3\\
           \end{pmatrix}=
           \begin{pmatrix}
      0 & -\cos\theta(t) & 0 \\
      \cos\theta(t) & 0 & -k\sin \theta(t) \\
      0 & k\sin\theta(t) & 0 \\
     \end{pmatrix}
           \begin{pmatrix}
           r_1\\
           r_2\\
           r_3\\
           \end{pmatrix}.
\end{equation}
 The problem of transferring
the system in Eq.~\eqref{eq:transfer} from
$(\cos\alpha,\sin\alpha,0, 0)^{T}$ to $(0, 0, \cos\beta,\sin\beta )^{T}$, 
reduces to finding $\theta(t)$, for steering the system from
 $(\cos\alpha,\sin\alpha,0)^T$ to $(0,\cos\beta,\sin\beta)^T$    
in minimal time in Eq.~\eqref{eq:main0}. We show that this is equivalent to finding the corresponding geodesic on the sphere, under the metric
\begin{equation}
\label{eq:metric}
g=\frac{k^2dr_1^2+dr_3^2}{k^2r_2^2}.
\end{equation}

By substituting for $\sin \theta(t)$ and $\cos \theta(t)$ from Eq.~\eqref{eq:main0},
the transfer time  $\tau=\int_{0}^\tau \sqrt{ [\sin \theta(t)]^2 + [\cos \theta(t)]^2}\ dt$
reduces to
$$ \tau=
\frac{1}{k}\int_0^\tau \sqrt{[k^2(\dot r_1)^2 + (\dot
r_3)^2]/r_2^2}\ dt = \frac{1}{k}\int_0^\tau L\ dt.
$$ Thus, minimizing $\tau$, amounts to
computing the geodesic under the metric $g$. The Euler-Lagrange
equations for the geodesic take the form
$$\frac{d}{dt}\left(\frac{\partial L}{\partial \dot{r}_1}\right) =
\left(\frac{\partial L}{\partial r_1}\right)\quad \text{ and }\quad
\frac{d}{dt}\left(\frac{\partial L}{\partial \dot{r}_3}\right) =
\left(\frac{\partial L}{\partial r_3}\right).$$ 
Note, $r_2^2=1-r_1^2-r_3^2$
and along the geodesic, $L=k$ is constant. We get, \be
\label{eq:Hamil}
\frac{d}{dt}\left(\frac{k^2\dot{r}_1}{r_2^2}\right)=L^2\frac{r_1}{r_2^2}\quad \text{ and }\quad \frac{d}{dt}\left(\frac{\dot{r}_3}{r_2^2}\right)=L^2\frac{r_3}{r_2^2},
\ee which implies that
\be
\label{eq:fderivative}
\frac{d}{dt}\left(\frac{k^2\dot{r}_1r_3-\dot{r}_3r_1}{r_2^2}\right)=(k^2-1)\frac{\dot{r}_1\dot{r}_3}{r_2^2}.
\ee Let $f=(k^2\dot{r}_1r_3-\dot{r}_3r_1)/r_2^2$. From
Eq.~\eqref{eq:Hamil}, we get 
\begin{equation*}
\frac{d}{dt}\left(\frac{k^2\dot{r}_1}{r_2}\right)= -f \frac{\dot{r}_3}{r_2}. 
\end{equation*}
Substitute with
$\dot{r}_1/r_2= -\cos\theta(t)$ and $\dot{r}_3/r_2=k \sin\theta(t)$,
and get
$$\frac{d}{dt}\left[-k^2\cos\theta(t)\right]=-f k \sin\theta(t),$$
so $\dot{\theta}=-f/k$. Differentiating again, we obtain from 
Eq.~\eqref{eq:fderivative}, that 
\begin{equation*}
 \ddot{\theta}=-\frac{1}{k}\frac{d}{dt}f
 =-\frac{k^2-1}{k}\frac{\dot{r}_1\dot{r}_3}{r_2^2}
 =(k^2-1)\cos\theta \sin\theta.
\end{equation*}
Using $a=k^2-1$,  we rewrite this as, 
\begin{equation}
\label{eq:elliptic}
\ddot{\theta} = \frac{a}{2} \sin 2 \theta(t). 
\end{equation}

The solution to Eq.~\eqref{eq:elliptic}, can be given in terms of 
an elliptic integral. Note that by multiplying both sides of Eq.~\eqref{eq:elliptic}, 
with $\dot{\theta}$, we get the equation of an ellipse in terms of the
coordinates $(\dot{\theta}, \cos[\theta])^{T}$ as
\begin{equation}\label{eq:c}
c = \dot{\theta}^2 (t) + a \cos^2 \theta(t), 
\end{equation}where $c$ is a constant.
Thus, we obtain that, 
$\dot{\theta} = \pm \sqrt{c - a \cos^2 \theta(t)}$ and
\begin{equation}\label{elliptic}
\int_{\theta(0)}^{\theta(t)} \frac{d \sigma}{\sqrt{c - a \cos^2 \sigma }}
= \pm t.
\end{equation} 
The left hand side of Eq.~\eqref{elliptic} is an elliptic integral \cite{BD:1993, WW:1963}. 
Equation~\eqref{eq:elliptic}, can be integrated if $\theta(0)$ and $\dot{\theta}(0)$
are both known explicitly. In transfers considered subsequently, only $\theta(0)$
is known and therefore one has to numerically search over the possible values of 
$\dot{\theta}(0)$, such that the resulting trajectory $X(t)$ achieves the desired transfer.
Note, guessing an initial value of $\dot{\theta}(0)$ is same as searching for the correct 
value of $c$ in Eq.~\eqref{eq:c}.

\section{Computation of Optimal Control Laws}
Now, consider the problem of steering the system of Eq.~\eqref{eq:main0}, from the initial
state $(r_1,r_2,r_3)^{T}=(1, 0, 0)^{T}$ to the target state $(0, \cos\beta, \sin\beta)^{T}$. 
For $\beta=\pi/2$, we get the target state $(0,0,1)^{T}$.
Recall that,
\begin{equation}\label{eq:f}
-k \dot{\theta} = f = - k^2 \cos \theta \frac{r_3}{r_2} - k \sin \theta \frac{r_1}{r_2}.
\end{equation}
At $t=0$, $r_1(0)= 1$ and $r_2(0)=r_3(0)= 0$. For $f$ to be finite, at $t=0$,
we should have $\sin \theta(0) = 0$. Therefore, in Eq.~\eqref{eq:elliptic}, 
the initial condition is $\theta(0)=0$. To solve Eq.~\eqref{eq:elliptic}, we
only need to know the initial value of $\dot{\theta}(0)$, which is the same as knowing 
the constant $c$ in Eq.~\eqref{eq:c}. Given $\theta(0)=0$ and the value of $\dot{\theta}(0)$, 
we can numerically solve first Eq.~\eqref{eq:elliptic} and then
Eq.~\eqref{eq:main0}. Consequently, we can determine for each value
of  $\dot{\theta}(0)$, the smallest time $s$ such that $r_{1}(s)=0$
by using a one-dimensional search, e.g., using the bisection
method (see pp.~46--51 of Ref.~\cite{BF:2005}). Thus, we can determine the 
value of $\dot{\theta}(0)$ such that $r_2(s)=\cos(\beta)$, again by using a one-dimensional search.
In summary, we have reduced the original optimzation problem to (combined) 
one-dimensional search problems.

\begin{figure}
\includegraphics{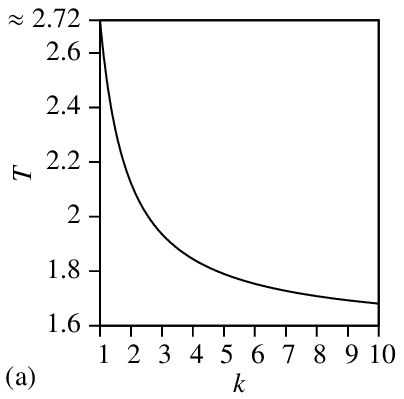}
\includegraphics{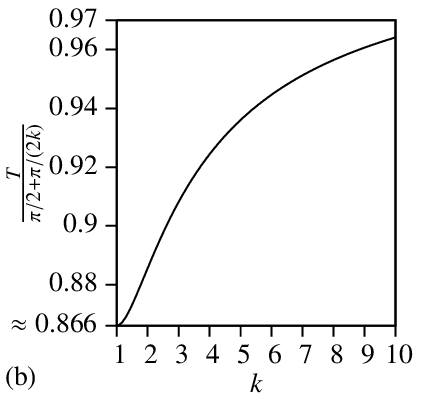}
\caption{For different $k$, we plot (a) the minimal time $T$ in units of
 $1/J_{12}$ and (b) the corresponding ratio $T/[\pi/2+\pi/(2k)]$.}
\label{fig:plot.time}
\end{figure}

\begin{figure}
\includegraphics{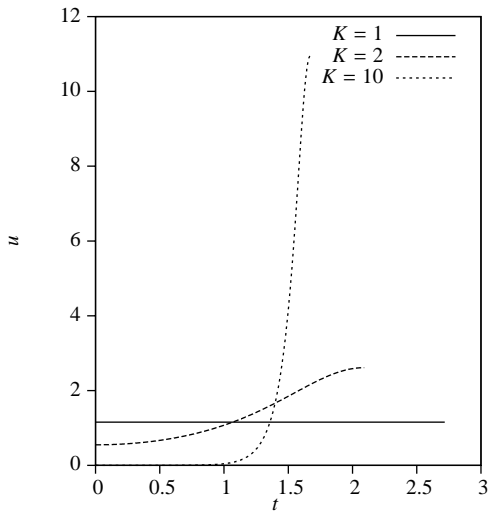}
\caption{The figure shows the shape of the optimal control $u(t)$ in Eq.~\eqref{eq:transfer}
for $k=1, 2, 10$. The time is in units of $1/J_{12}$. The control amplitude is in units of $J_{12}$.
Rightmost datapoint of each control $u(t)$ is not displayed for reasons of numerical accuracy.}
\label{fig:plot.control}
\end{figure}

For different values of $k$, we plot the minimal time $T=T(k)$
to transfer $(r_1,r_2,r_3)^{T}=(1, 0, 0)^{T}$ to $(0, 0, 1)^{T}$ in
Fig.~\ref{fig:plot.time}(a) in units of $1/J_{12}$.
For example, when $k=1$, i.e., $J_{12}=J_{23}=J$, it takes 
$2.72/J$ amount of time, which is about 
$86.6\%$ of $\pi/J $, the time needed using the conventional method. 
Figure~\ref{fig:plot.time}(b), shows the ratio of the the minimal time $T(k)$ 
to the time $\pi/2+\pi/(2k)$ obtained using the conventional strategy. We define this ratio $\eta(k)$. 
In both the plots
$k \geq 1$ is considered, as the case for $k < 1$ can be derived from this.
 Figure~\ref{fig:plot.control}, shows the optimal control 
$u(t)$ in Eq.~\eqref{eq:transfer} for different values of $k$.

Observe that $T(1/k) = k T(k)$. Let $u(t, k)$ be the time optimal
control for steering the system of Eq.~\eqref{eq:transfer} 
from $(1, 0, 0, 0)^{T}$ to $(0, 0, 0, 1)^{T}$. Then the 
control $v(t) = u(T-t, k)$, will steer the same system from $(0, 0, 0, 1)^{T}$ 
to $(1, 0, 0, 0)^{T}$ in the same time, which is also minimal for this 
transfer. 
Let $Y = (x_4, x_3, x_2, x_1)^{T}$ and consider the 
control $v(t) = u(T-t, k)$. 
Then, we have 
$$\frac{dY}{d \tau} = k B(1/k, v)Y. $$ But, we have just remarked 
that the minimal time to steer $Y$ from $(1, 0, 0, 0)^{T}$ to $(0, 0, 0, 1)^{T}$ is $T(k)$, 
and it follows that $T(1/k)/k = T(k)$.  
It also follows that the optimal control 
$u(t, 1/k) = u[T(k)-t/k , k]$, where $t \in [0, k T(k)]$. Note $\eta(k) = \eta(1/k)$.

\begin{figure}
\includegraphics{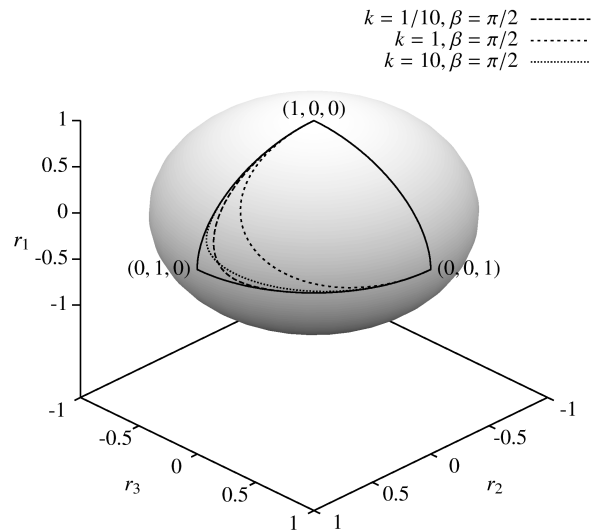}
\caption{Geodesics 
to transfer from $(r_{1},r_{2},r_{3})^{T}=(1,0,0)^{T}$ to $(0,0,1)^{T}$,
where $k=1/10$, $k=1$, and $k=10$.}
\label{fig:main3}
\end{figure}

{\it Remark 1.} In Fig.~\ref{fig:main3}, we present the geodesics 
of the metric of Eq.~\eqref{eq:metric}
for $\beta = \pi/2$ and $k\in\{1/10,1,10\}$, where 
$\beta = \tan^{-1}[r_3(T)/r_2(T)]$, i.e., $r_1(T)=r_2(T)=0$ and $r_3(T)=1$.
Observe that the geodesics for the case $k > 1$, bend more towards the point
$(r_1, r_2, r_3) = (0, 1, 0)$, before approaching the final point $(0, 0, 1)$.
In an intuitive way to understand this, consider first the limit $k\gg 1$. In this limit, 
the minimum time to steer Eq.~\eqref{eq:main0} from $(1, 0, 0)$ to $(0, 0, 1)$ is essentially
the time required to steer the system (of Eq.~\eqref{eq:main0}) to the equator. This is achieved 
fastest by moving from $(1, 0, 0)$, directly to $(0, 1, 0)$ and keeping $\theta(0)=0$. 
As $k$ is decreased to $1$, the geodesics gradually move away from $(0, 1, 0)$, implying 
$\dot{\theta}(0)$ for $k>1$ is smaller than for the case $k=1$.
In the special case of $k=1$, Eq.~\eqref{eq:elliptic} reduces to 
$\dot{\theta} = C$, a constant. This case was been studied in detail 
in \cite{KHSY:2007, YGK:2007}. Therefore, in the numerical computation of optimal control,
the search for true $\dot{\theta}(0)$ can be restricted to the interval $[0, C]$.

\begin{figure}
\includegraphics{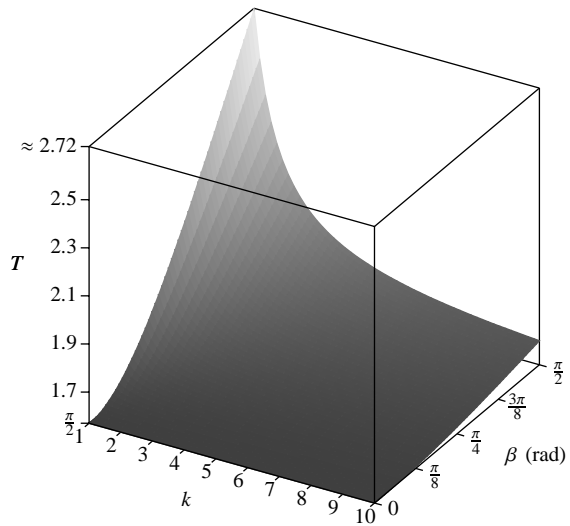}
\caption{For different $k$ and $\beta=\tan^{-1}[r_3(T)/r_2(T)]$, we plot the minimal 
time $T$ in units of $1/J_{12}$}
\label{fig:time}
\end{figure}

\begin{figure}
\includegraphics{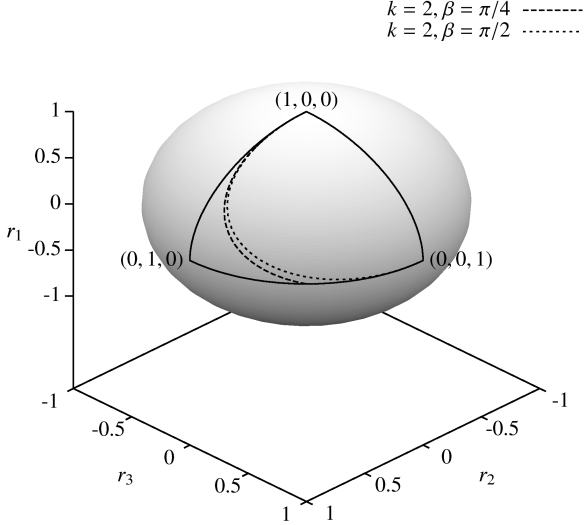}
\caption{Geodesics 
to transfer from $(r_{1},r_{2},r_{3})^{T}=(1,0,0)^{T}$ to $(0, \cos\beta,\sin\beta)^{T}$,
where $k=2$ and $\beta\in\{\pi/4,\pi/2\}$.}
\label{fig:main2}
\end{figure}

In Fig.~\ref{fig:time}, we plot the minimal time $T=T(k,\beta)$
to transfer $(r_1,r_2,r_3)^{T}=(1, 0, 0)^{T}$ to $(0, \cos\beta, \sin\beta)^{T}$,
for different values of $k$ and $\beta = \tan^{-1}[r_3(T)/r_2(T)]$.
Figure~\ref{fig:main2}, shows the geodesics of the metric of Eq.~\eqref{eq:metric}
for $k=2$, for two different values $\beta\in\{\pi/4,\pi/2\}$
of the terminal point $(0, \cos\beta, \sin\beta)^{T}$.

\begin{figure}
\includegraphics{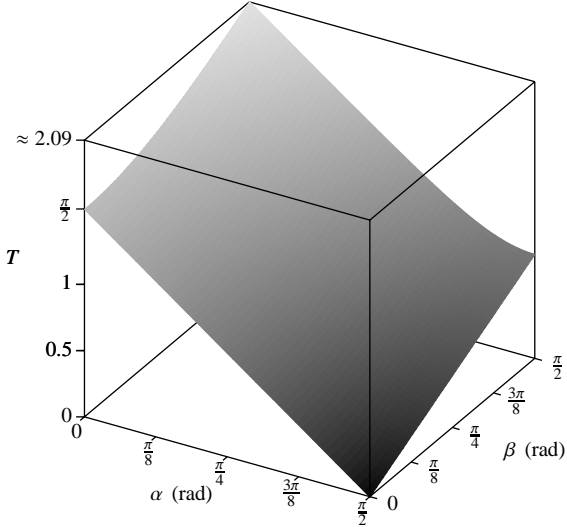}
\caption{For $k=2$ and different values of $\alpha$ and $\beta$, we plot the minimal 
time $T$ in units of $1/J_{12}$}
\label{fig:time_K2}
\end{figure}

If we transfer the system from $(x_1,x_2,x_3,x_4)^{T}=(\cos\alpha,\sin\alpha,0, 0)^{T}$ to $(0, 0, \cos\beta,\sin\beta )^{T}$,
where $\alpha >0$,  we do not know the value of $\theta(0)$. But
from Eq.~\eqref{eq:f}, we have the  relationship $\dot{\theta}(0)=\sin[\theta(0)] \cot(\alpha)$
between $\dot{\theta}(0)$ and $\theta(0)$. Thus, we can apply similar 
search methods for the right initial conditions 
as in the case of $\alpha=0$. 
For $k=2$ and different values of $\alpha$ and $\beta$, we plot
in Fig.~\ref{fig:time_K2} the minimal time $T=T(\alpha,\beta)$
to transfer $(r_1,r_2,r_3)^{T}=(\cos\alpha, \sin\alpha, 0)^{T}$ to $(0, \cos\beta, \sin\beta)^{T}$.

\section{Efficient Creation of Coherences in Ising Spin Chains with Unequal Couplings}
We now consider a generalization of the problem treated above. 
We consider a linear chain of $n$ spins, placed in a static
external magnetic field in the $z$-direction, with unequal Ising type
couplings between next neighbors \cite{Isi:1925, Cas:1989}. In a
suitably chosen (multiple) rotating frame, which rotates with each
spin at its resonant frequency, the free evolution of the spin system 
is given by the coupling Hamiltonian,
$$H_c=2 \sum_{l=1}^{n-1} J_{l, l+1} I_{lz}I_{(l+1)z}. $$
As before, we assume that the Larmor frequency of the spins are well separated 
compared to the coupling constant $J_{l,l+1}$, so that 
we can selectively rotate each spin at a rate much faster than the 
evolution of the couplings. 

We first consider the problem of synthesizing a unitary transformation
$U$, which efficiently transfers a single spin coherence
represented by the initial operator $I_{1x}$ to a 
multiple spin order represented by the operator 
$2^{n-1}\prod_{l=1}^{n}I_{lz}$. 

The conventional strategy for achieving this transfer is achieved through 
the following stages
\begin{eqnarray*}
\aligned &I_{1x}\stackrel{H_c}\rightarrow2I_{1y}I_{2z}\stackrel{I_{2y}}\rightarrow
2I_{1y}I_{2x}\stackrel{H_c}\rightarrow 4I_{1y}I_{2y}I_{3z}\\
&\stackrel{I_{3y}}\rightarrow4I_{1y}I_{2y}I_{3x}\rightarrow
\cdots\rightarrow 2^{n-1}(\prod_{l=1}^{n-1}I_{ly})I_{nz}
\endaligned
\end{eqnarray*}
where each evolution represents an appropriate evolution by rotation angle $\pi/2$. 
The final state $2^{n-1}(\prod_{l=1}^{n-1}I_{ly})I_{nz}$, 
is locally equivalent to $2^{n-1}\prod_{l=1}^{n}I_{lz}$. The whole transfer 
involves $n-1$ evolution steps under the natural Hamiltonian, where the first step 
takes only $\pi/(2J_{l, l+1})$ units of time, resulting in a total time of 
$\sum_{l=1}^{n-1} \pi/(2J_{l, l+1})$.
We now formulate the problem of this transfer as a problem of optimal
control and derive time efficient strategies for achieving this transfer.

To simplify notation, we introduce the following symbols for the
expectation values of operators that play a key part in the transfer:
\begin{gather*}
x_1=\langle I_{1x}\rangle,
x_2=\langle 2I_{1y}I_{2z}\rangle,\\
x_3=\langle 2I_{1y}I_{2x}\rangle,
x_4=\langle 4I_{1y}I_{2y}I_{3z}\rangle,
\ldots,\\
x_{2n-3}= \langle 2^{n-1}I_{1y}I_{2y}I_{3y}\cdots I_{(n-1)x} \rangle,\\
\text{and } x_{2n-2}=\langle 2^{n}I_{1y}I_{2y}I_{3y}\cdots I_{(n-1)y}I_{nz}\rangle.
\end{gather*}

Let $X = (x_1, x_2, x_3, \dots, x_{2n-2})^T$.
Expressing the time in units of 
$1/J_{12}$, the evolution of the system is
given by
\begin{equation}
\label{eq:transfer2} \frac{dX}{dt}= 
    \begin{pmatrix}
      0 & -k_1 & 0 & 0 & 0 & 0 & \cdots\\
      k_1 & 0 & -u_1 & 0 & 0 & 0 &\cdots\\
      0 & u_1 & 0 & -k_2 & 0 & 0 &\cdots\\
      0 & 0 & k_2 & 0 & -u_2 & 0 &\cdots\\
      0 & 0 & 0 & u_2 & 0 & -k_3 &\cdots\\
      0 & 0 & 0 & 0 & k_3 & 0   & \cdots\\
      \vdots & \vdots & \vdots & \vdots & \vdots& \vdots & \ddots
\end{pmatrix}X,
\end{equation}
where
$u_l$ are the control parameters 
representing the amplitude of the 
$y$ pulse on spin $l+1$ and $k_l = J_{l, l+1}/J_{12}$.
The problem now is to find the optimal
$u_l(t)$, steering the system from 
$(1,0,0,\cdots,0)^{T}$ to $(0,0,\cdots,0,1)^{T}$ in 
the minimal time.

We divide the optimal transfer problem into multiple steps. Let $l = 1, \dots, n-1$. 
Consider the step 
that steers $(x_{2l-1}, x_{2l}, x_{2l+1}, x_{2l+2})^{T}$
from the initial state $(\cos \beta_{l}, \sin \beta_{l}, 0, 0)^{T}$ to the target state
$(0, 0, \cos \beta_{l+1}, \sin \beta_{l+1})^{T}$, by optimal choice of $u_l$.
\begin{equation}
\label{eq:transfer3} \frac{d}{dt}\begin{pmatrix}
           x_{2l-1}\\
           x_{2l}\\
           x_{2l+1}\\
           x_{2l+2}
           \end{pmatrix}=
    \begin{pmatrix}
      0 & -k_l & 0 & 0\\
      k_{l} & 0 & -u_l & 0\\
      0 & u_l & 0 & -k_{l+1}\\
      0 & 0 & k_{l+1} & 0
    \end{pmatrix}
    \begin{pmatrix}
           x_{2l-1}\\
           x_{2l}\\
           x_{2l+1}\\
           x_{2l+2}
    \end{pmatrix},
\end{equation}

We denote the minimal time for this transfer as $T(\beta_{l}, \beta_{l+1}, k_{l}, k_{l+1})$, as it depends on
the parameters, $(\beta_{l}, \beta_{l+1}, k_l, k_{l+1})$. Note that 
$$ k_l T(\beta_{l}, \beta_{l+1}, k_l, k_{l+1}) = T(\beta_{l}, \beta_{l+1}, 1, k_{l+1}/k_l). $$

We can then break the original transfer in Eq.~\eqref{eq:transfer2}, 
into sub-transfers denoted by 
$$ \beta_{1} \rightarrow \beta_{2} \dots \rightarrow  \beta_{l} \dots \rightarrow
\beta_{n-1}, $$ where both $\beta_{1} =0$ and $\beta_{n-1}=\pi/2$. The control
$u_l$ in Eq.~\eqref{eq:transfer3} should be so chosen that the time for transfer
$\beta_{l}$ to $\beta_{l+1}$ is minimized. Furthermore, the choice of
intermediate $\beta_{l}$ should be made to minimize the total time of transfer
$$ J_1(0) = \sum_{l=1}^{n-2} T(\beta_{l}, \beta_{l+1}, k_l, k_{l+1}). $$

The notation for $J_1(0)$ denotes that the index $l$ in the above summation begins 
with $l=1$ with initial angle $\beta_{1}=0$. Now suppose, we have computed
$J_{l+1}(\xi)$ for all $\xi \in [0, \pi/2 ]$. Then we can compute
$J_l(\beta_{l})$ from the equation below as
\begin{equation}
\label{eq:dynamic}
J_l(\beta_{l}) = \min_{\eta} \{ J_{l+1}(\eta) +  T(\beta_{l}, \eta, k_l, k_{l+1}) \}.
\end{equation}

We can solve Eq.~\eqref{eq:dynamic} iteratively starting from $l=n-2$ and proceeding backwards and
remembering that 
\begin{equation*}
J_{n-2}(\beta_{n-2}) = T(\beta_{n-2}, \pi/2, k_{n-2}, k_{n-1}).
\end{equation*}
Note $T(\beta_{l}, \eta, k_l, k_{l+1})$ in Eq.~\eqref{eq:dynamic} can be explicitly computed,
as we showed in the previous section. Therefore Eq.~\eqref{eq:dynamic} helps us to 
compute the optimal $\eta$ and the corresponding optimal control $u_l$.
 
Now, the problem is reduced to a simple dynamic programming problem, which can be 
solved efficiently. 

\begin{figure}
\includegraphics{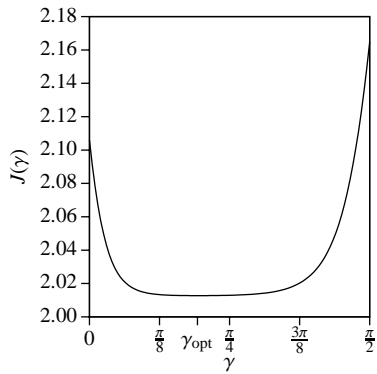}
\caption{The figure shows the plot of function $J(\gamma)$ in Eq.~\eqref{eq:time1}, 
as a function of $\gamma$.}
\label{fig:special3k_100}
\end{figure}

{\it Example 2.} We consider as an example, an Ising spin chain with four spins
with $J_{12}/(2 \pi) = 91$ Hz, $J_{23}/(2 \pi) = 15$ Hz and 
$J_{34}/(2 \pi) = 55$ Hz. This system represents the popular \cite{CFPS:1996}
HNCACO experiment in multidimensional NMR, where
first spin is the proton, the second one represent $^{15}N$, the third and fourth one are $^{13}C$.
For the transfer in Eq.~\eqref{eq:transfer2}, the dynamic programming equations for the 
minimum time is in 
units of $1/J_{23}$, can be written as $\min_{\gamma} J(\gamma)$, where
\begin{equation}
\label{eq:time1}
J(\gamma) = \{ T(0, \gamma, k_1, 1) +  T(\gamma, \frac{\pi}{2}, 1, k_2) \}
\end{equation}
where $k_1 = J_{12}/J_{23}$ and $k_2 = J_{34}/J_{23}$. Figure~\ref{fig:special3k_100}
shows the plot of $J(\gamma)$ and the minimum is achieved at $\gamma_{\text{opt}} \approx .193 \pi$ and 
the corresponding minimum value of $J(\gamma_{\text{opt}})$ is $\approx 2.01$. The conventional transfer strategy will take $\pi(1 + 1/k_1 + 1/k_2)/2 \approx 2.26$, units of time, which is $\approx 12.2\%$ longer than
the proposed efficient methodology.

\section{Conclusion}
In this article, we studied the problem of efficient creation of multiple spin 
order in an Ising spin chain with unequal couplings. We first analyzed in detail the 
system of three linearly coupled spins. We showed 
that the time optimal pulse sequences for creating multiple spin order in this system
can be obtained by computing geodesics on a sphere under a special metric
and that the solution to the resulting Euler Lagrange equation is 
related to the solution of a nonlinear oscillator equation 
$\ddot{\theta} = A \sin \theta $. We showed that the minimum times and optimal control 
laws for the studied problem can be explicitly computed and provide significant gains over 
conventional methods. The methods developed in the three spin case were then exploited to find efficient strategies for manipulating the dynamics of Ising spin chains with unequal couplings. 
It is expected that these methods will find immediate applications in 
coherent spectroscopy and quantum information processing.


\bibliography{spinchain}

\begin{thebibliography}{28}
\expandafter\ifx\csname natexlab\endcsname\relax\def\natexlab#1{#1}\fi
\expandafter\ifx\csname bibnamefont\endcsname\relax
  \def\bibnamefont#1{#1}\fi
\expandafter\ifx\csname bibfnamefont\endcsname\relax
  \def\bibfnamefont#1{#1}\fi
\expandafter\ifx\csname citenamefont\endcsname\relax
  \def\citenamefont#1{#1}\fi
\expandafter\ifx\csname url\endcsname\relax
  \def\url#1{\texttt{#1}}\fi
\expandafter\ifx\csname urlprefix\endcsname\relax\def\urlprefix{URL }\fi
\providecommand{\bibinfo}[2]{#2}
\providecommand{\eprint}[2][]{\url{#2}}

\bibitem[{\citenamefont{Khaneja et~al.}(2001)\citenamefont{Khaneja, Brockett,
  and Glaser}}]{KBG:2001}
\bibinfo{author}{\bibfnamefont{N.}~\bibnamefont{Khaneja}},
  \bibinfo{author}{\bibfnamefont{R.}~\bibnamefont{Brockett}}, \bibnamefont{and}
  \bibinfo{author}{\bibfnamefont{S.~J.} \bibnamefont{Glaser}},
  \bibinfo{journal}{Phys. Rev. A} \textbf{\bibinfo{volume}{63}},
  \bibinfo{pages}{032308} (\bibinfo{year}{2001}).

\bibitem[{\citenamefont{Bennett et~al.}(2002)\citenamefont{Bennett, Cirac,
  Leifer, Leung, Linden, Popescu, and Vidal}}]{BCL:2002}
\bibinfo{author}{\bibfnamefont{C.~H.} \bibnamefont{Bennett}},
  \bibinfo{author}{\bibfnamefont{J.~I.} \bibnamefont{Cirac}},
  \bibinfo{author}{\bibfnamefont{M.~S.} \bibnamefont{Leifer}},
  \bibinfo{author}{\bibfnamefont{D.~W.} \bibnamefont{Leung}},
  \bibinfo{author}{\bibfnamefont{N.}~\bibnamefont{Linden}},
  \bibinfo{author}{\bibfnamefont{S.}~\bibnamefont{Popescu}}, \bibnamefont{and}
  \bibinfo{author}{\bibfnamefont{G.}~\bibnamefont{Vidal}},
  \bibinfo{journal}{Phys. Rev. A} \textbf{\bibinfo{volume}{66}},
  \bibinfo{pages}{012305} (\bibinfo{year}{2002}).

\bibitem[{\citenamefont{Vidal et~al.}(2002)\citenamefont{Vidal, Hammerer, and
  Cirac}}]{VHC:2002}
\bibinfo{author}{\bibfnamefont{G.}~\bibnamefont{Vidal}},
  \bibinfo{author}{\bibfnamefont{K.}~\bibnamefont{Hammerer}}, \bibnamefont{and}
  \bibinfo{author}{\bibfnamefont{J.~I.} \bibnamefont{Cirac}},
  \bibinfo{journal}{Phys. Rev. Lett.} \textbf{\bibinfo{volume}{88}},
  \bibinfo{pages}{237902} (\bibinfo{year}{2002}).

\bibitem[{\citenamefont{Hammerer et~al.}(2002)\citenamefont{Hammerer, Vidal,
  and Cirac}}]{HVC:2002}
\bibinfo{author}{\bibfnamefont{K.}~\bibnamefont{Hammerer}},
  \bibinfo{author}{\bibfnamefont{G.}~\bibnamefont{Vidal}}, \bibnamefont{and}
  \bibinfo{author}{\bibfnamefont{J.~I.} \bibnamefont{Cirac}},
  \bibinfo{journal}{Phys. Rev. A} \textbf{\bibinfo{volume}{66}},
  \bibinfo{pages}{062321} (\bibinfo{year}{2002}).

\bibitem[{\citenamefont{Reiss et~al.}(2002)\citenamefont{Reiss, Khaneja, and
  Glaser}}]{RKG:2002}
\bibinfo{author}{\bibfnamefont{T.~O.} \bibnamefont{Reiss}},
  \bibinfo{author}{\bibfnamefont{N.}~\bibnamefont{Khaneja}}, \bibnamefont{and}
  \bibinfo{author}{\bibfnamefont{S.~J.} \bibnamefont{Glaser}},
  \bibinfo{journal}{J. Magn. Reson.} \textbf{\bibinfo{volume}{154}},
  \bibinfo{pages}{192} (\bibinfo{year}{2002}).

\bibitem[{\citenamefont{Zeier et~al.}(2004)\citenamefont{Zeier, Grassl, and
  Beth}}]{ZGB:2004}
\bibinfo{author}{\bibfnamefont{R.}~\bibnamefont{Zeier}},
  \bibinfo{author}{\bibfnamefont{M.}~\bibnamefont{Grassl}}, \bibnamefont{and}
  \bibinfo{author}{\bibfnamefont{T.}~\bibnamefont{Beth}},
  \bibinfo{journal}{Phys. Rev. A} \textbf{\bibinfo{volume}{70}},
  \bibinfo{pages}{032319} (\bibinfo{year}{2004}).

\bibitem[{\citenamefont{Yuan and Khaneja}(2005)}]{YK:2005}
\bibinfo{author}{\bibfnamefont{H.}~\bibnamefont{Yuan}} \bibnamefont{and}
  \bibinfo{author}{\bibfnamefont{N.}~\bibnamefont{Khaneja}},
  \bibinfo{journal}{Phys. Rev. A} \textbf{\bibinfo{volume}{72}},
  \bibinfo{pages}{040301(R)} (\bibinfo{year}{2005}).

\bibitem[{\citenamefont{Yuan}(2006)}]{Yua:2006}
\bibinfo{author}{\bibfnamefont{H.}~\bibnamefont{Yuan}}, Ph.D. thesis,
  \bibinfo{school}{Harvard University} (\bibinfo{year}{2006}).

\bibitem[{\citenamefont{Zeier}(2006)}]{Zei:2006}
\bibinfo{author}{\bibfnamefont{R.~M.} \bibnamefont{Zeier}},
  \emph{\bibinfo{title}{Lie-theoretischer Zugang zur Erzeugung unit{\"a}rer
  Transformationen auf Quantenrechnern}}
  (\bibinfo{publisher}{Universit{\"a}tsverlag Karlsruhe},
  \bibinfo{address}{Karlsruhe}, \bibinfo{year}{2006}), \bibinfo{note}{{P}h.D.\
  thesis, Universit{\"a}t Karlsruhe, 2006}.

\bibitem[{\citenamefont{Khaneja et~al.}(2002)\citenamefont{Khaneja, Glaser, and
  Brockett}}]{KGB:2002}
\bibinfo{author}{\bibfnamefont{N.}~\bibnamefont{Khaneja}},
  \bibinfo{author}{\bibfnamefont{S.~J.} \bibnamefont{Glaser}},
  \bibnamefont{and} \bibinfo{author}{\bibfnamefont{R.}~\bibnamefont{Brockett}},
  \bibinfo{journal}{Phys. Rev. A} \textbf{\bibinfo{volume}{65}},
  \bibinfo{pages}{032301} (\bibinfo{year}{2002}).

\bibitem[{\citenamefont{Reiss et~al.}(2003)\citenamefont{Reiss, Khaneja, and
  Glaser}}]{RKG:2003}
\bibinfo{author}{\bibfnamefont{T.~O.} \bibnamefont{Reiss}},
  \bibinfo{author}{\bibfnamefont{N.}~\bibnamefont{Khaneja}}, \bibnamefont{and}
  \bibinfo{author}{\bibfnamefont{S.~J.} \bibnamefont{Glaser}},
  \bibinfo{journal}{J. Magn. Reson.} \textbf{\bibinfo{volume}{165}},
  \bibinfo{pages}{95} (\bibinfo{year}{2003}).

\bibitem[{\citenamefont{Khaneja et~al.}(2007)\citenamefont{Khaneja, Heitmann,
  Sp{\"o}rl, Yuan, Schulte-Herbr{\"u}ggen, and Glaser}}]{KHSY:2007}
\bibinfo{author}{\bibfnamefont{N.}~\bibnamefont{Khaneja}},
  \bibinfo{author}{\bibfnamefont{B.}~\bibnamefont{Heitmann}},
  \bibinfo{author}{\bibfnamefont{A.}~\bibnamefont{Sp{\"o}rl}},
  \bibinfo{author}{\bibfnamefont{H.}~\bibnamefont{Yuan}},
  \bibinfo{author}{\bibfnamefont{T.}~\bibnamefont{Schulte-Herbr{\"u}ggen}},
  \bibnamefont{and} \bibinfo{author}{\bibfnamefont{S.~J.}
  \bibnamefont{Glaser}}, \bibinfo{journal}{Phys. Rev. A}
  \textbf{\bibinfo{volume}{75}}, \bibinfo{pages}{012322}
  (\bibinfo{year}{2007}).

\bibitem[{\citenamefont{Khaneja and Glaser}(2002)}]{KG:2002}
\bibinfo{author}{\bibfnamefont{N.}~\bibnamefont{Khaneja}} \bibnamefont{and}
  \bibinfo{author}{\bibfnamefont{S.~J.} \bibnamefont{Glaser}},
  \bibinfo{journal}{Phys. Rev. A} \textbf{\bibinfo{volume}{66}},
  \bibinfo{pages}{060301(R)} (\bibinfo{year}{2002}).

\bibitem[{\citenamefont{Yuan et~al.}(2007)\citenamefont{Yuan, Glaser, and
  Khaneja}}]{YGK:2007}
\bibinfo{author}{\bibfnamefont{H.}~\bibnamefont{Yuan}},
  \bibinfo{author}{\bibfnamefont{S.~J.} \bibnamefont{Glaser}},
  \bibnamefont{and} \bibinfo{author}{\bibfnamefont{N.}~\bibnamefont{Khaneja}},
  \bibinfo{journal}{Phys. Rev. A} \textbf{\bibinfo{volume}{76}},
  \bibinfo{pages}{012316} (\bibinfo{year}{2007}).

\bibitem[{\citenamefont{Ernst et~al.}(1997)\citenamefont{Ernst, Bodenhausen,
  and Wokaun}}]{EBW:1997}
\bibinfo{author}{\bibfnamefont{R.~R.} \bibnamefont{Ernst}},
  \bibinfo{author}{\bibfnamefont{G.}~\bibnamefont{Bodenhausen}},
  \bibnamefont{and} \bibinfo{author}{\bibfnamefont{A.}~\bibnamefont{Wokaun}},
  \emph{\bibinfo{title}{Principles of Nuclear Magnetic Resonance in One and Two
  Dimensions}} (\bibinfo{publisher}{Clarendon Press},
  \bibinfo{address}{Oxford}, \bibinfo{year}{1997}), \bibinfo{note}{reprinted
  with corrections}.

\bibitem[{\citenamefont{Yuan and Khaneja}(2006)}]{YK:2006}
\bibinfo{author}{\bibfnamefont{H.}~\bibnamefont{Yuan}} \bibnamefont{and}
  \bibinfo{author}{\bibfnamefont{N.}~\bibnamefont{Khaneja}}, in
  \emph{\bibinfo{booktitle}{Proceedings of the 45th IEEE Conference on Decision
  and Control, 2006}} (\bibinfo{address}{San Diego, USA},
  \bibinfo{year}{2006}), pp. \bibinfo{pages}{3117--3120}.

\bibitem[{\citenamefont{Nielsen
  et~al.}(2006{\natexlab{a}})\citenamefont{Nielsen, Dowling, Gu, and
  Doherty}}]{NDGD:2006}
\bibinfo{author}{\bibfnamefont{M.~A.} \bibnamefont{Nielsen}},
  \bibinfo{author}{\bibfnamefont{M.~R.} \bibnamefont{Dowling}},
  \bibinfo{author}{\bibfnamefont{M.}~\bibnamefont{Gu}}, \bibnamefont{and}
  \bibinfo{author}{\bibfnamefont{A.~C.} \bibnamefont{Doherty}},
  \bibinfo{journal}{Science} \textbf{\bibinfo{volume}{311}},
  \bibinfo{pages}{1133} (\bibinfo{year}{2006}{\natexlab{a}}).

\bibitem[{\citenamefont{Nielsen}(2006)}]{Nie:2006}
\bibinfo{author}{\bibfnamefont{M.~A.} \bibnamefont{Nielsen}},
  \bibinfo{journal}{Quantum Inf. Comput.} \textbf{\bibinfo{volume}{6}},
  \bibinfo{pages}{213} (\bibinfo{year}{2006}).

\bibitem[{\citenamefont{Nielsen
  et~al.}(2006{\natexlab{b}})\citenamefont{Nielsen, Dowling, Gu, and
  Doherty}}]{NDGD:2006b}
\bibinfo{author}{\bibfnamefont{M.~A.} \bibnamefont{Nielsen}},
  \bibinfo{author}{\bibfnamefont{M.~R.} \bibnamefont{Dowling}},
  \bibinfo{author}{\bibfnamefont{M.}~\bibnamefont{Gu}}, \bibnamefont{and}
  \bibinfo{author}{\bibfnamefont{A.~C.} \bibnamefont{Doherty}},
  \bibinfo{journal}{Phys. Rev. A} \textbf{\bibinfo{volume}{73}},
  \bibinfo{pages}{062323} (\bibinfo{year}{2006}{\natexlab{b}}).

\bibitem[{\citenamefont{Brockett}(1982)}]{Bro:1982}
\bibinfo{author}{\bibfnamefont{R.~W.} \bibnamefont{Brockett}}, in
  \emph{\bibinfo{booktitle}{New Directions in Applied Mathematics}}, edited by
  \bibinfo{editor}{\bibfnamefont{P.~J.} \bibnamefont{Hilton}} \bibnamefont{and}
  \bibinfo{editor}{\bibfnamefont{G.~S.} \bibnamefont{Young}}
  (\bibinfo{publisher}{Springer}, \bibinfo{address}{New York},
  \bibinfo{year}{1982}), pp. \bibinfo{pages}{11--27}.

\bibitem[{\citenamefont{Baillieul}(1975)}]{Bal:1975}
\bibinfo{author}{\bibfnamefont{J.~B.} \bibnamefont{Baillieul}}, Ph.D. thesis,
  \bibinfo{school}{Harvard University} (\bibinfo{year}{1975}).

\bibitem[{\citenamefont{Montgomery}(2002)}]{Mon:2002}
\bibinfo{author}{\bibfnamefont{R.}~\bibnamefont{Montgomery}},
  \emph{\bibinfo{title}{A Tour of Subriemannian Geometries, Their Geodesics and
  Applications}}, no.~\bibinfo{number}{91} in \bibinfo{series}{Mathematical
  surveys and monographs} (\bibinfo{publisher}{American Mathematical Society},
  \bibinfo{address}{Providence}, \bibinfo{year}{2002}).

\bibitem[{\citenamefont{Ising}(1925)}]{Isi:1925}
\bibinfo{author}{\bibfnamefont{E.}~\bibnamefont{Ising}}, \bibinfo{journal}{Z.
  Physik} \textbf{\bibinfo{volume}{31}}, \bibinfo{pages}{253}
  (\bibinfo{year}{1925}).

\bibitem[{\citenamefont{Caspers}(1989)}]{Cas:1989}
\bibinfo{author}{\bibfnamefont{W.~J.} \bibnamefont{Caspers}},
  \emph{\bibinfo{title}{Spin systems}} (\bibinfo{publisher}{World Scientific},
  \bibinfo{address}{Singapore}, \bibinfo{year}{1989}).

\bibitem[{\citenamefont{Brockett and Dai}(1993)}]{BD:1993}
\bibinfo{author}{\bibfnamefont{R.~W.} \bibnamefont{Brockett}} \bibnamefont{and}
  \bibinfo{author}{\bibfnamefont{L.}~\bibnamefont{Dai}}, in
  \emph{\bibinfo{booktitle}{Nonholonomic Motion Planning}}, edited by
  \bibinfo{editor}{\bibfnamefont{Z.}~\bibnamefont{Li}} \bibnamefont{and}
  \bibinfo{editor}{\bibfnamefont{J.~F.} \bibnamefont{Canny}}
  (\bibinfo{publisher}{Kluwer Academic Publishers}, \bibinfo{year}{1993}), pp.
  \bibinfo{pages}{1--21}.

\bibitem[{\citenamefont{Whittaker and Watson}(1963)}]{WW:1963}
\bibinfo{author}{\bibfnamefont{E.~T.} \bibnamefont{Whittaker}}
  \bibnamefont{and} \bibinfo{author}{\bibfnamefont{G.~N.}
  \bibnamefont{Watson}}, \emph{\bibinfo{title}{A course of modern analysis}}
  (\bibinfo{publisher}{Cambridge University Press},
  \bibinfo{address}{Cambridge}, \bibinfo{year}{1963}), \bibinfo{edition}{4th}
  ed.

\bibitem[{\citenamefont{Burden and Faires}(2005)}]{BF:2005}
\bibinfo{author}{\bibfnamefont{R.~L.} \bibnamefont{Burden}} \bibnamefont{and}
  \bibinfo{author}{\bibfnamefont{J.~D.} \bibnamefont{Faires}},
  \emph{\bibinfo{title}{Numerical Analysis}} (\bibinfo{publisher}{Thomson
  Brooks/Cole}, \bibinfo{address}{Belmont}, \bibinfo{year}{2005}),
  \bibinfo{edition}{8th} ed.

\bibitem[{\citenamefont{Cavanagh et~al.}(1996)\citenamefont{Cavanagh,
  Fairbrother, Palmer, and Skelton}}]{CFPS:1996}
\bibinfo{author}{\bibfnamefont{J.}~\bibnamefont{Cavanagh}},
  \bibinfo{author}{\bibfnamefont{W.~J.} \bibnamefont{Fairbrother}},
  \bibinfo{author}{\bibfnamefont{A.~G.} \bibnamefont{Palmer}},
  \bibnamefont{and} \bibinfo{author}{\bibfnamefont{N.~J.}
  \bibnamefont{Skelton}}, \emph{\bibinfo{title}{Protein NMR Spectroscopy:
  Principles and Practice}} (\bibinfo{publisher}{Academic Press},
  \bibinfo{address}{San Diego}, \bibinfo{year}{1996}).

\end{thebibliography}

\end{document}